# Scaling in stratocumulus fields: an emergent property

Tianle Yuan[1,2]
[1]Laboratory for Climate and Radiation, NASA Goddard Space Flight Center
[2]Joint Center for Earth Technology Systems, University of Maryland, Baltimore County, MD, 20740

Abtract

Marine stratocumulus clouds play a critical role in the Earth's climate system. They display a wide range of complex behavior at many different spatiotemporal scales. Precipitation in these clouds is usually light, but it is vital for their systematic changes and organization state. Here we show, through analysis of observational data, that stratocumulus clouds across the globe has a universal scaling in the size distribution of potentially precipitating areas. Such scaling is argued to be an emergent property of the cloud system. It results from numerous interactions at the microscopic scales.

Text

Marine stratocumulus clouds (Sc) occupy large areas of the Earth's ocean surface[1]. They exert strong radiative cooling on the climate by reflecting much more solar energy than the underlying ocean[2]. Determination of how Sc responds to aerosols and climate change is of critical importance because small changes in Sc coverage can either strongly enhance or partially compensate greenhouse gas warming[3,4]. Recent decades have seen progress in process-level understanding of Sc through analysis of data from field campaigns, remote sensing and modeling[5-9]. Yet, representing these fine-scale processes in climate models is still a challenge[10] and large uncertainties remains as demonstrated by inter-model comparison studies[11].

Despite the stratiform appearance Sc fields are composed of numerous small convective eddies driven by longwave cooling at the top. A web of non-linear and intricately coupled meteorological, oceanic, radiative and microphysical processes repeatedly interacts across a range of scales to drive Sc evolution[8]. These nonlinear and interrelated interactions can render small perturbations highly consequential for the overall system behavior under certain conditions. The coupled system also gives rise to intriguing organization patterns such as closed and open cellular convection[12]. Particularly puzzling is the sudden appearance of pockets of open cells (POCs) within an otherwise overcast cloud deck [6,13,14]. Some POCs can punctuate a whole Sc deck and reduce cloud fraction by more than 50% over millions of square miles in a short period of time[15], rendering them critical for Sc cloud fraction, while others remain confined in size for a long time (SOM). On the other hand, simple statistical relationships exist between macroscopic cloud properties and the mean state despite fine-scale intricacies[16]. Given these observations here we suggest that a complex system view[17] can prove advantageous to grasping aspects of overall Sc system behavior.

A complex system is a system that is composed of an ensemble of interacting elements whose collective behavior cannot be trivially derived from understanding of individual elements. It provides a fitting perspective to understand organization of Sc fields: we can think of individual convective eddy, which is principally well understood, as the basic interacting element and macro-scale organization as a collective behavior. One critical step when studying such a system is to identify certain macroscopic phenomenon and to characterize its fluctuations in size. Here we are interested in macroscopic organization of precipitation (or drizzling) in Sc fields. Precipitation is intimately related to Sc organization and evolution [6,13,15,18]. For example, both modeling and observational studies show precipitation is necessary for the appearance of POCs[15,19-21]. Precipitation is also instrumental in maintaining the open cellular convection[22].

We indentify a potentially precipitating pouch (simply as 'pouch' latter) as the important macroscopic object. It is defined as a group of interconnected (4-neighbor connectivity) cloudy areas whose cloud liquid water path ($L = \int_{cloud\_base}^{cloud\_top} \rho q_l dz$, $\rho$, $q_l$ and $z$ are air density, water mixing ratio and height, respectively) is no less than 180 gm$^{-2}$[23]. Pouches are organized structures imbedded within seemingly homogeneous Sc decks (Figure 1)[24]. and they are areas susceptible to POC formations given their high $L$ and the primary dependence of Sc precipitation on $L$ [25] (see demonstration in SOM). They are the only breeding ground of POC formation because of the necessary condition of precipitation[19,20,26].

Cloud optical depth ($\tau$) and droplet effective radius ($r_{eff}$) from MODIS[27] are used to derive $L$ with $L = \frac{5}{9}\rho_w r_{eff} \tau$ ($\rho_w$ is the liquid water density), assuming an adiabatic cloud [28], at 1km resolution. $L$ is also available from AMSR-E at 13km resolution[23]. An automated program scans each MODIS granule and AMSR-E swath to find individual pouches. For every pouch we record its size ($s$, simply the number of pixels in an individual pouch), mean ($\bar{L}$), and standard deviation ($\delta_L$) of $L$, the geographical location and perimeter length ($p$) among others (for more details see SOM). We analyze data for three major Sc regions, off coasts of California, Peru and Namibia, to explore geographic and climatic variability. These three regions have their own distinct large-scale, oceanographic and microphysical regimes despite similar general conditions that favor Sc formation. To explore seasonal and interannual variability we select two months, June and September in both 2003 and 2008, for each region that have contrasting cloud fraction and $\bar{L}$. Two MODIS instruments, one at mid-morning and the other at early afternoon, partially capture the diurnal variability. We note that our results are not sensitive to the threshold of $L$ when it is increased by up to 20%.

Spatial distribution of pouch occurrence frequency ($f_p$) and its relation to spatial variability of cloud fraction illustrate the importance of pouches in Sc evolution (Figure 2). We include only pouches that are large than 100 pixels because the typical size of a POC or an open cell exceeds 100 km$^2$ in area [12,286,13,14]. $f_p$ increases rapidly from the coast towards remote ocean and peaks around 90°W and 18°S (Figure X) and further downwind it starts to quickly decrease (Figure X). The area of fast reduction is exactly where open cell organization strongly increases their occurrence and closed cell organization transforms into open cell[28]. This again strongly suggests that statistically large pouches play a pivotal role in POC formation and closed-to-open transition because they are likely the only breeding grounds to open up Sc decks. This closed-to-open transition quickly decreases Sc cloud fraction, demonstrating a powerful impact of cloud organization on evolution of cloud fraction[28] (Figure X and SOM). For comparison, pouch frequency increases with the mean flow for areas near the coast (between 5S and 25S) and correspondingly cloud fraction remains very high. The internal properties of pouches also undergo dramatic changes within the transition area such as the strong increase in pouch $\overline{L}$ and $\delta_L$. In the following we show that their statistical organizations can offer illuminating clues to macroscopic organization of Sc systems.

First, the area-perimeter ($s-p$) relation is used to characterize the geometric structure of pouches. The $s-p$ relation follows the formula $p \propto \sqrt{s}^D$ as shown in Figure 3a. This formula applies smoothly without any break for 4 orders of magnitude in spatial scales suggesting the absence of characteristic scale within the range studied here. The exponent $D$ characterizes the degree of contortion of the perimeter. Here $D$ takes a value of 1.22 (Figure 3a) that is noticeably smaller than that of deep convective cloud precipitation[29], but comparable to warm cumulus clouds[30]. This indicates that pouches are smoother in appearance than convective rain and clouds. The value of $D$ has implications for underlying processes. In the Kolmogorov turbulence model $D$ has values of 4/3 for isobars, which agrees well with 1.35 for deep convective clouds and rain in real atmosphere[29]. The smaller $D$ here implies that in the shallow boundary layer processes such as latent heating, radiation and entrainment significantly impact the organization of pouches. The ability of precipitation to organize Sc at the mesoscale may be particularly important to understanding the difference in

$D^{28,31}$. Finally, we highlight an extraordinary invariability of $D$. It takes the same value of about 1.22 no matter which region, season, year or time of the day we choose to analyze. The insensitivity of $D$ to these environmental variations indicates possible universality in the nature of pouch formation and organization processes.

Figure 3b shows normalized ($\int_{min}^{max} P(s)ds = 1$) probability, $P(s)$, density distributions of pouch size for three regions. The distributions include all pouches sampled during September 2003. All three distributions can be characterized by a power-law, $P(s) \propto s^{-\alpha}$. The existence of a simple power-law distribution that describes pouches across the whole spectrum (spanning four decades here) reflects a stunning unity of the underlying processes: the large pouches are described in the same way as the smallest ones. Furthermore, the power-law distribution is the hallmark of scale-invariance and self-similarity, which is also demonstrated by the $s - p$ relationship above (Figure 3a). We use data from AMSR-E to further illustrate the scale invariance since this instrument samples nearly the same cloud fields as the afternoon MODIS but with much coarser resolution. This resolution difference offers a new view of Sc fields under a different magnifying glass. With the 'coarse-grained' AMSR-E data pouch size distribution still follows a power-law with nearly identical $\alpha$: pouches essentially look the same under different magnifying glasses (Figure 3c). The scale-invariance also means that there is no typical Sc pouch scale within the scale span analyzed here.

The degree of similarity among different Sc regions ($\alpha = 2 \pm 0.02$) is truly remarkable, considering the large geographic variations. For example, the Californian Sc deck is in an opposite phase of seasonal cycle than the two southern hemisphere Sc regions[16]. The atmosphere stability and cloud fraction are close to maximum in September for two southern hemisphere regions while they are in transition from maximum to minimum for the Californian deck [16]. Moreover, there exist strong contrasts in critical variables such as droplet number concentration, drizzle frequency, $\bar{L}$ and sea surface temperature among three Sc regions[32,33]. The similarity therefore implies that the web of non-linear and coupled processes repeatedly interacts to effectively smooth out detailed differences and give rise to a pouch organization that is insensitive to details. This is the essence of the idea of emergence in complex systems: generation of new higher-level system properties from repeated lower-level interactions that cannot be trivially reducible to the constituent parts.

Further analyses reveal universality (in the sense that the same power-law applies regardless of the season, time of the day, year or geographic region) of this emergent behavior. Figure 3c shows pouch sizes distributions for June 2003 to illustrate their sensitivity (or the lack of) to seasonal variation. They again follow a power-law with, strikingly, the same exponents ($\alpha = 2 \pm 0.02$). For individual Sc regions $\alpha$ stays constant regardless of the seasonal changes of a host of environmental conditions. $\alpha$ is also insensitive to inter-annual variations. Meanwhile, the absolute number of pouches does have strong diurnal, geographical, seasonal and interannual variations. The constant $\alpha$ values suggest that pouches fluctuate, responding to variations in environmental conditions, in such an orderly way, out of almost infinite possibilities, that relative change for each pouch size bin is the same.

Diurnal changes provide clues to the underlying mechanisms for this orderly fluctuation. Pouches in both the afternoon and the morning follow the same power distribution as shown in figure X. The number of pouches, however, is reduced in the afternoon across the map (Figure X and X). This is due mainly to strong solar heating that decreases cloud $\bar{L}$ (mean $L$ for all cloudy pixels) because the synoptic and large-scale forcings are nearly identical for the two sampling times. Pouch $\bar{L}$ (mean $L$ for pouch pixels), to the contrary, increases in the afternoon despite the overall decrease of cloud $\bar{L}$ and number of pouches except in the lower-left corner of the map where pouch $\bar{L}$ peaks. The relatively small afternoon decrease in cloud overall fraction prohibits the possibility of pouches disappearing altogether as result of solar heating. This possibility seems also too bipolar, i.e. pouches either remaining unchanged or disappearing altogether under the same diurnal change, to be viable. Visual inspections also rule out occurrence of POCs, which concentrates during nighttime[15], as a viable explanation.

We instead propose that a simple splitting mechanism can explain the observed insensitivity of $\alpha$. Edge pixels (those with $L$ close to 180 gm$^{-2}$) in a pouch can become non-pouch pixels due to solar heating and related mesoscale organization. Transitioning of these edge pixels splits a large pouch into two or more smaller ones. As the splitting process applies to pouches across the size spectrum, the small pouches increase their number, but many disappear in our samples due to either sensor resolution limit or reduction of $\bar{L}$, explaining the overall decrease in number of pouches. The splitting mechanism is also consistent with the slight increase in pouch $\bar{L}$ in the afternoon: the pouches in a sense protect themselves by preferentially shedding edge pixels while maintaining the moister centers[18]. Most critically, power-law distributions are robust against splitting of pouches (SOM). These factors make the extremely simple process of splitting a viable mechanism. The opposite mechanism, i.e. merging, can therefore be invoked to understand scenarios when cloud $\bar{L}$ increases. Indeed, if we abstract the collective effect of numerous process-level interactions as simply a splitting/merging process, a stochastic network model can correctly simulate the observed power-law pouch size distribution[34]. This simple explanatory model is also consistent with the observed insensitivity of size distribution to environmental conditions because it is abstract by construct and is completely unaware of changes in $\bar{L}$ or other parameters.

Not only is this abstraction viable, observations presented in this study really demand such abstraction. For example, pouch organizations (Figures 2 and 3) are so regular over such a range of scales that they are likely to be independent of microscopic and macroscopic details, making abstract models imperative as well as viable. The universality of the pouch organization lends further support of abstraction at higher-level and usage of simple explanatory models. Such abstract, explanatory models offer means to hide details while retaining the essence of the higher-level organization, which makes it suitable to studying emergent behavior within the complex Sc system. With appropriate abstraction and simplification, aspects of dynamics in a complex system can be investigated with dramatically reduced degrees of freedom[35].

Our analyses reveal the scale-invariant nature of pouch organization, which demonstrates the fundamental challenge of accounting for Sc in large-scale models: every scale is equally important. This challenge is best manifested by the slow progress at which coarse-resolution models are improving to capture mean state and variability of macroscopic Sc properties[36]. Indeed, models with improved resolution and physics display dramatic improvement[37]. However, emergent phenomena such as POC formation are still poorly represented[37]. Analyses and model studies of macroscopic variability and organization of clouds may be critical for removing this particular model inadequacy[28,38].

To this end we find emergence of macroscopic organization out of repeated local interactions, taking a complex system view. Similar emergent behavior can be found in a host of seemingly unrelated systems such as biology, ecology and even human society in addition to physical sciences[17]. The similarity is arguably rooted in the existence of hierarchy structure at different levels, which can be abstracted in stochastic models regardless of the exact nature of interactions. The existence of hierarchy provides a potentially fertile avenue for representing Sc (and other cloud[34]) fields in large-scale models. Knowledge and tools from the fledging field of complex system research can be borrowed to stochastically represent macroscopic cloud behaviors under different large-scale environments[39].

The greatest challenge for applying network theory tools would be to appropriately abstract collective effect of numerous lower-low interactions. Observational analyses and process-oriented modeling (i.e. reductionism) can facilitate progress in this direction. In particular, we argue that satellite observations provide the necessary data depth and breadth to facilitate system minded analyses. Such observations can provide critical metrics for assessing high-resolution process-oriented model performance. It is also at this junction that parallel lines of system-based[34,35,40] approach and reductionism may merge to best complement each other: Reductionism is necessary and useful to understand the behavior of individual interacting element while system-based models capture the macroscopic collective behavior. We thus believe combination of these two approaches will expedite our understanding of Sc and other clouds in general.

To summarize, the potentially precipitating pouches are breeding grounds for POC formation and play important role in Sc evolution. Taking a complex system view we show emergent macroscopic behaviors of pouches. Universal scaling of pouch sizes and area-perimeter relationship is resilient to a variety of environmental conditions, which can be understood with a simple network model. A complex system approach might be particularly beneficial for representing macroscopic properties of Sc in large-scale models.

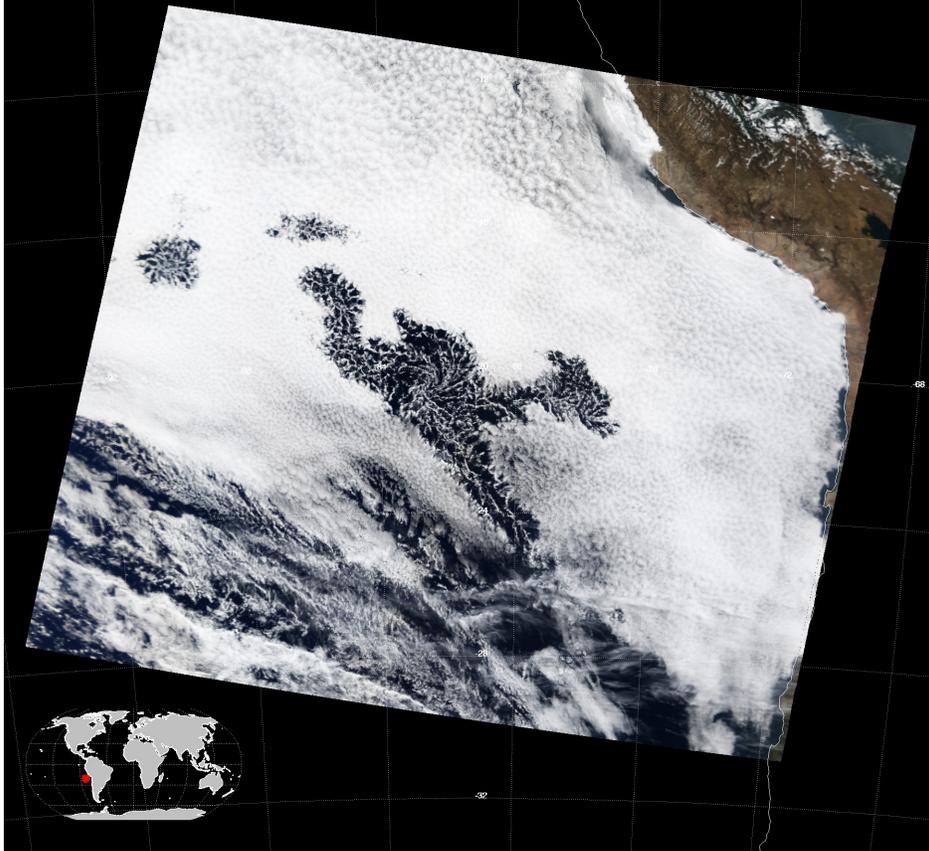

Figure 1: example of pockets of open cells breaking up a stratocumulus deck. The image is taken by MODIS over southeast Pacific in 2003.

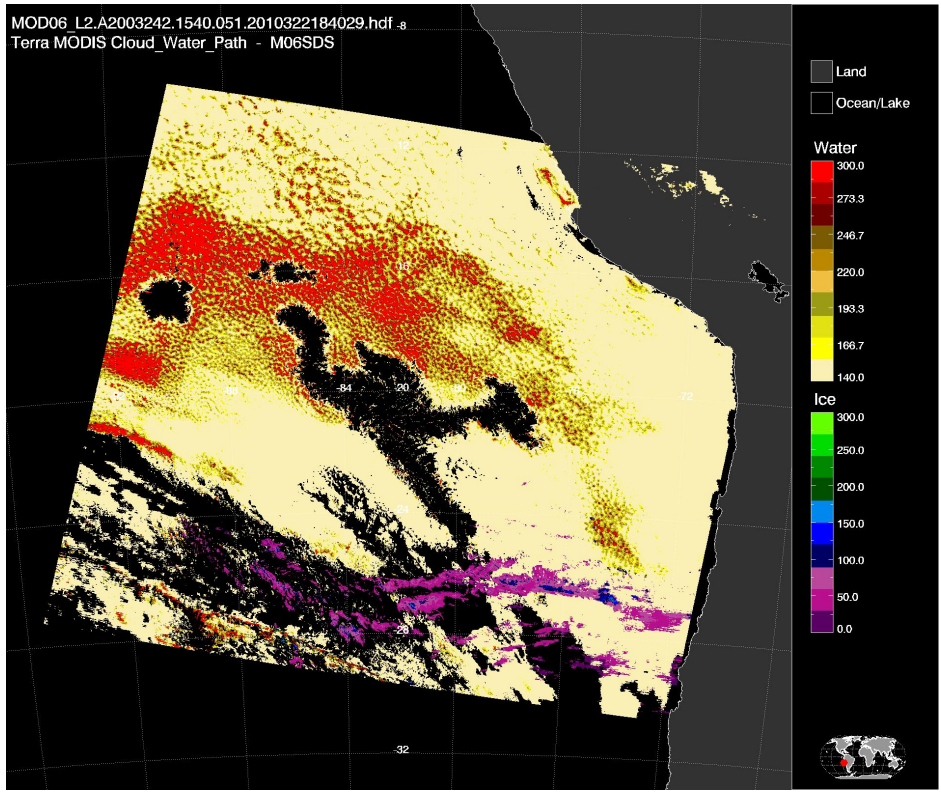

Figure 1b: MODIS water path (both liquid and ice) for the same scene. Warm (cold) clouds are for warm (cold) clouds. L can be obtained by multiple values in the figure by 0.83. Note the 'running dog' shaped POC spans two distinct regions: the head, neck and the front leg are embedded within a low L region while the rest in a high L region (see SOM).

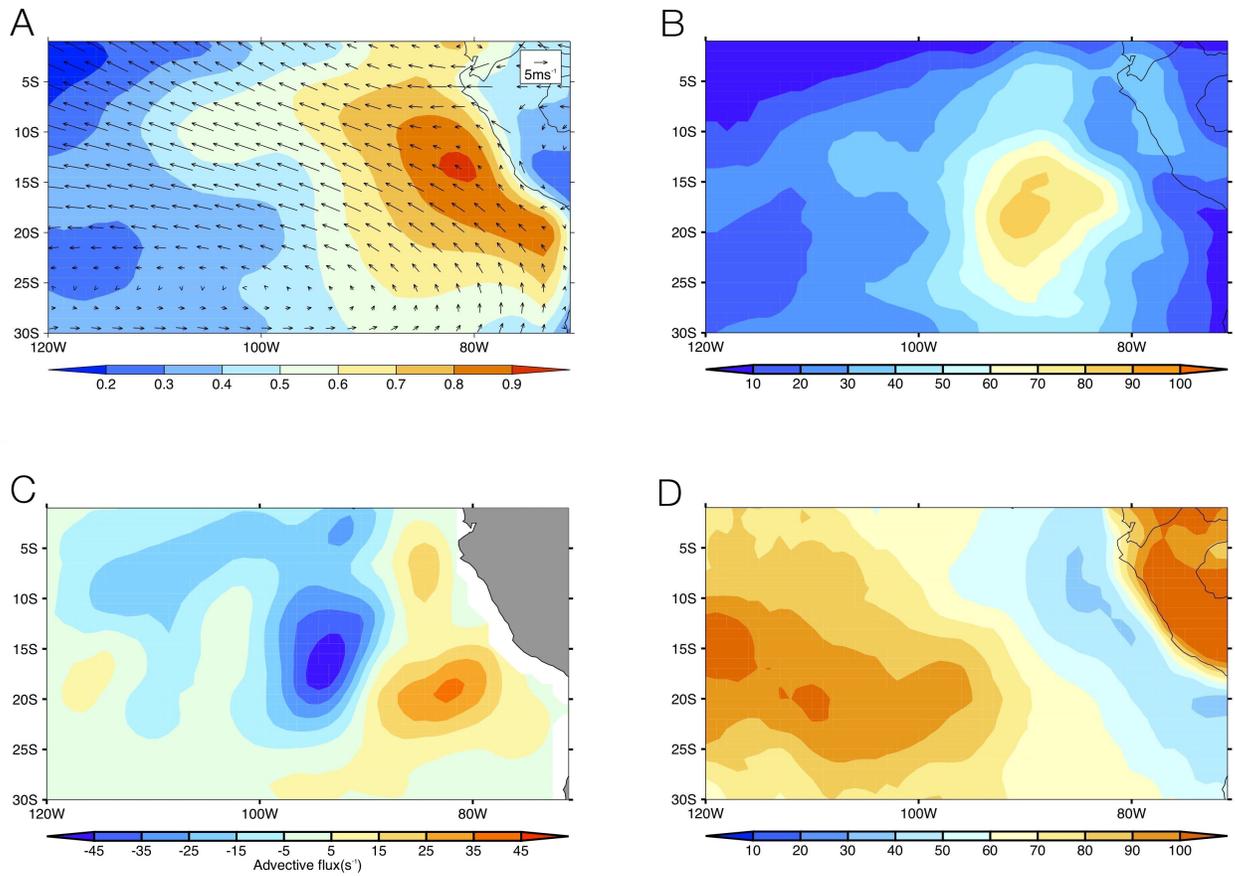

Figure 2: A) MODIS liquid cloud fraction with wind vectors from MERRA overlaid B) number of large pouches C) Pouch number advection flux calculated based on monthly mean wind and pouch frequency distribution (panel B) D) Standard deviation of $L$ for large pouches.

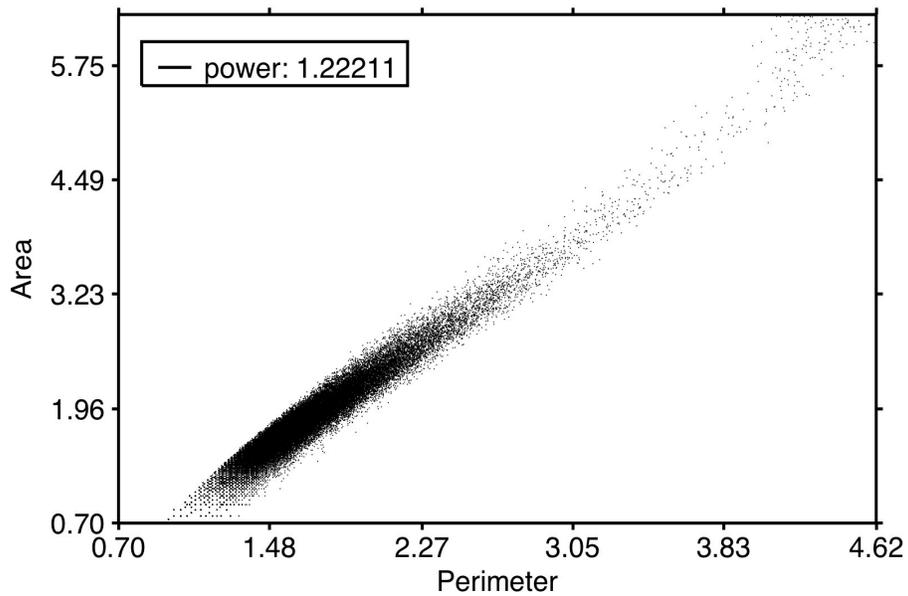

Figure 3a: the area-perimeter diagram plotted on log-log scale.

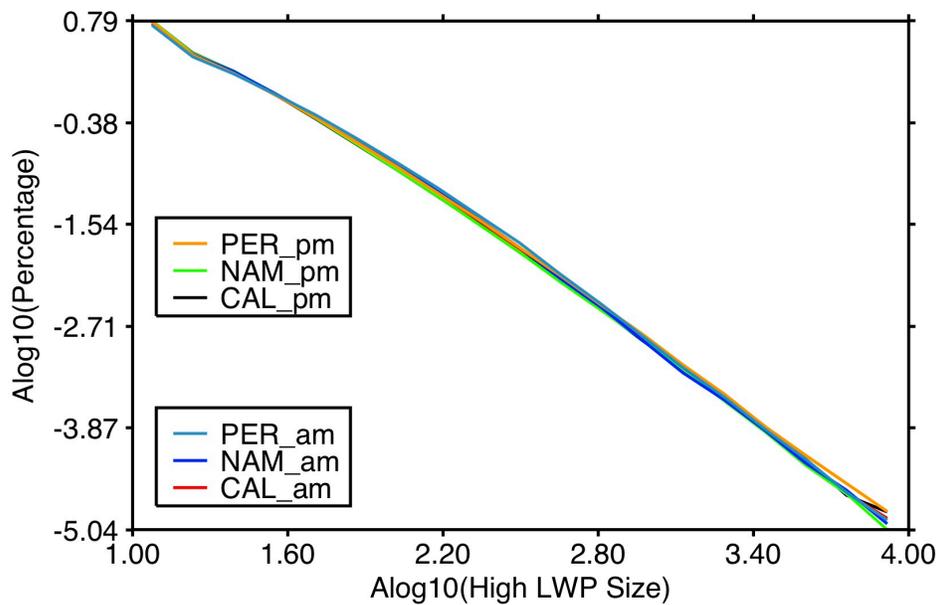

Figure 3b: MODIS pouch size distributions for three Sc regions using both Terra and Aqua data in September, 2003.

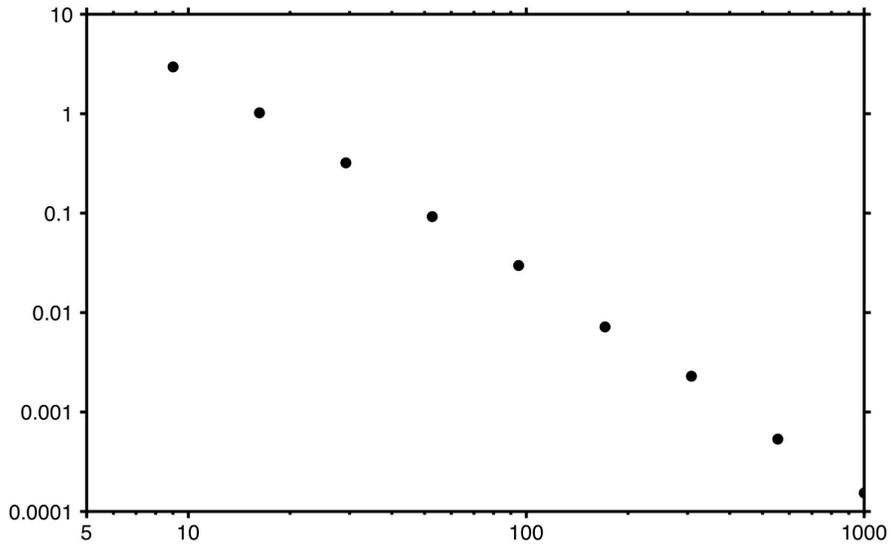

Figure 3c: AMSR pouch size distribution. The real size of the pouches is approximately 169 times of the number in x-axis.

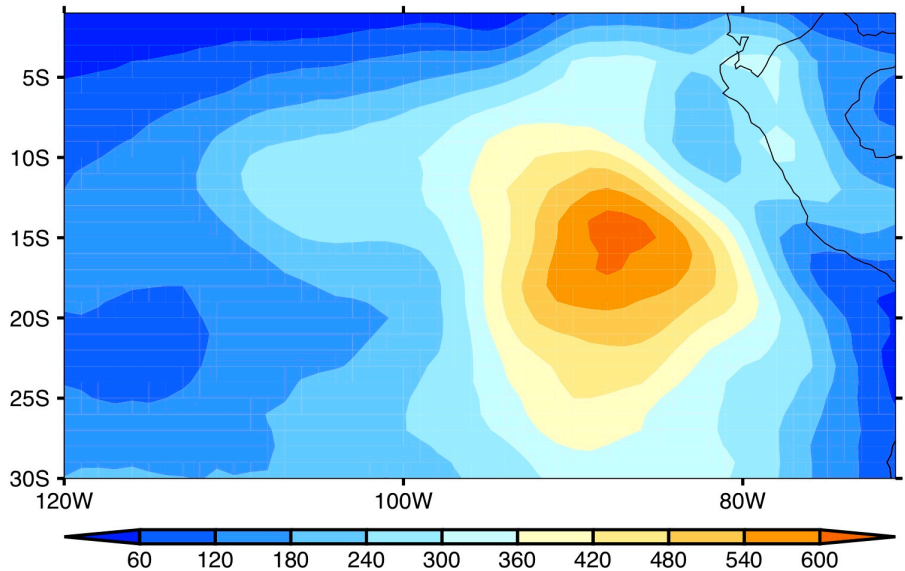

Figure 4a total number of pouches that have size less than 10000. This map is for Terra MODIS.

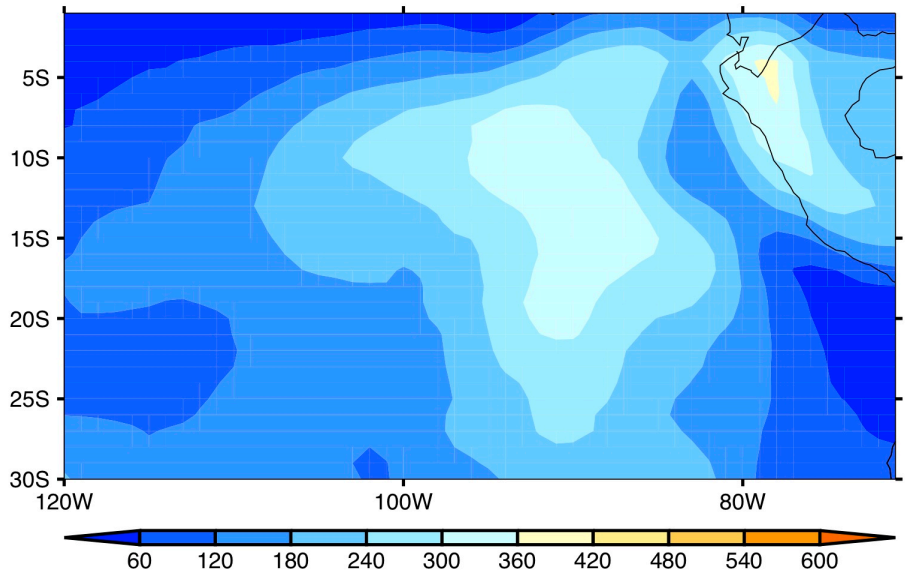

Figure 4b: same as 4a but for Aqua MODIS.

Figure 4c: see Figure 3a. A different plot will be made here. (to be added).